\documentclass [a4paper, 12pt]{article}
\usepackage{psfrag}
\usepackage{epsfig}
\usepackage{graphicx}
\usepackage{epstopdf}
\usepackage{amssymb}
\begin{document}
\date{}
\author{Mrinal Kanti Debnath, Soumyajit Pramanick$^1$,
Sudeshna DasGupta$^2$\\
~and 
Nababrata Ghoshal
\footnote{Corresponding author. E-mail: ghoshaln@gmail.com
}\\ 
Department of Physics, Ramsaday College,\\
Amta, Howrah, West Bengal, INDIA\\
$^1$Department of Physics, St John College,\\
Dimapur 797112, INDIA\\
$^2$Department of Physics, Lady Brabourne College,\\
Kolkata 700017, INDIA\\
$^\ast$Department of Physics,\\
Ramsaday College, Howrah, West Bengal, INDIA \\
}
\title {
	Phase diagram of a biaxial nematogenic lattice model: A Monte Carlo simulation study
 } 
 
\maketitle

\begin{abstract}
The phase diagram for a lattice system of biaxial molecules possessing $D_{2h}$ symmetry and interacting with the Straley's quadrupolar pair potential in Sonnet-Virga-Durand parameterization [A. M. Sonnet, E. G. Virga, and G. E. Durand, Phys. Rev. E {\bf67}, 061701 (2003)] has been determined using Monte Carlo simulation. Our results confirm that the nematogenic model yields both the uniaxial and biaxial nematic macroscopic phases along with a tricritical point in the transition from uniaxial to biaxial nematics as has been predicted in mean field theory. By analyzing the behaviour of a free-energy-like function, derived from the probability distributions of energy, the order of phase transitions are detected. A conclusive numerical evidence in support of the existence of a tricritical point on the uniaxial-biaxial transition line in the phase diagram is reported.  
Although the nature of the phase diagram is qualitatively identical as obtained in the mean field study however the location of triple point differs significantly from theoretical prediction.   
 \end{abstract}

\section{INTRODUCTION}
\indent Over the last three decades great interest has been paid to
thermotropic biaxial nematic liquid crystals \cite{sluc}, whose
existence was first predicted by Freiser \cite{fre} in 1970 from
mean-field molecular theory. Freiser showed that deviations from assumed cylindrical symmetry of nematogenic molecules should result in the formation of a biaxial nematic phase apart from the conventional uniaxial nematic phase. This study predicts a first-order transition from an isotropic ($I$) to a uniaxial nematic phase ($N_U$) at a higher temperature 
and a second-order transition from a uniaxial nematic to a biaxial nematic phase ($N_B$) at a lower temperature. 
In his model, Freiser assumed the shape
of the mesogenic molecules to be lath-like having $D_{2h}$ symmetry. Shortly afterwards, a number of
theoretical \cite{alb,str,luc1} and computer simulation  studies \cite{luc2,allen,bis} on single-component models consisting of non-cylindrical molecules having $D_{2h}$ symmetry have been performed. These studies have generated 
phase diagrams consisting
of three distinct macroscopic phases namely, the
isotropic phase ($I$), the uniaxial nematic phase ($N_U$)
and the biaxial nematic phase ($N_B$). A direct second-order $I$ - $N_B$ transition is predicted as well at a particular molecular geometry called the self-dual point \cite{chi, gho1}. 

\indent On the experimental front, there have
been a number of reports \cite{mal1,mal2,cha1,cha2} of observations
of a thermotropic $N_B$ phase since 1986. However, none of these claims proved to be correct \cite{luc3}. 
Recently
there have been claims of the identification of the thermotropic
$N_B$ phase for bent-core ("banana-shaped") molecules \cite{mad,ach,luc4}
and for organosiloxane tetrapode molecules \cite{mer,fig}. These new findings have fuelled further investigations \cite{bat,gho2,gho3} on phase biaxiality in thermotropic nematics. 

\indent Recently, a mean-field (MF) model within the general expression of Straley's quadrupolar potential for biaxial (board-like) molecules has been proposed \cite{son}. 
This model predicts a tricritical point ($i.e.$, the point at which the transition changes from first to second order) on the $N_U$ - $N_B$ transition line and a triple point, where the three phases ($i.e.$, $I$, $N_U$, and $N_B$) coexist in the phase diagram in biaxiality-temperature plane. 
Subsequently another mean field study \cite{mat1} has shown that the same model predicts another tricritical point on the line of the direct $I$-$N_B$ transition. 

\indent In the mean time some Monte Carlo (MC) simulations \cite{rom1,rom2,mat2} for the same pair potential, called as the Sonnet-Virga-Durand (SVD) parameterization \cite{sluc2}, are performed which 
confirm qualitatively the predictions of the above MF studies. Simultaneously, a qualitatively different phase diagram has been reported by Latha $\textit{et al.}$ \cite{latha1,latha2}. In their MC simulations they employed an entropic sampling technique and observed that the direct $I$ - $N_B$ transition, as has been predicted by previous MF studies, is replaced beyond a certain value of the biaxiality parameter by an additional intermediate biaxial nematic phase.  

\indent The existence of the tricritical point on the $N_U$ - $N_B$ transition line has also been confirmed in their experimental investigation by Merkel $\textit{et al.}$ on two liquid crystalline systems of organo-siloxane tetrapodes \cite{mer}. 

\indent However, a detailed numerical study of the theoretically predicted and  experimentally detected tricritical point in the transition from uniaxial to biaxial nematics is not available as yet. In this communication, we present the phase diagram in  biaxiality-temperature plane generated from the results of a Monte Carlo simulation in a simple cubic lattice system where biaxial molecules possesing $D_{2h}$ symmetry interact with  nearest neighbours via the Straley's quadrupolar pair potential in SVD parameterization. In particular, apart from the usual thermodynamic observables, a free-energy-like function, derived from the energy probability distributions, has been used to characterize the order of phase transitions and to confirm the existence of the tricritical point on the $N_U$ - $N_B$ transition line. Another significant finding of our investigation is that the first-order uniaxial-biaxial transition is much weaker than the usual weak first-order isotropic-uniaxial transition.  
Our study, although, confirms the qualitative nature of the phase diagram obtained from the mean-field calculation, provides new results for the coordinates of triple point in the phase diagram. 

The plan of this paper is as follows: in Sec. II we discuss
the Straley's quadrupolar potential in SVD parameterization and the technical details of the simulations; in Sec. III we present the results.
Conclusions are presented in Sec. IV.

\section{THE MODEL AND SIMULATIONS}
\indent Here we consider a lattice model of biaxial prolate molecules
possessing $D_{2h}$ symmetry (board-like), whose centers of mass are
attached with a simple-cubic lattice. For a board-like molecule, we associate three orthogonal unit vectors,
$\{\mathbf{e}, \mathbf{e}_\perp, \mathbf{m}\}$, which are the eigenvectors of any polarizability tensor of 
the mesogenic molecule. Among these, $\mathbf{m}$ represents the long molecular symmetry axis, 
while $\mathbf{e}$ and $\mathbf{e}_\perp$ are the two short transverse axes.

\indent We use the Straley's pair potential in SVD parameterization between two identical neighboring
molecules, say the  $i^{th}$ and $j^{th}$ molecules,
\begin{equation}\label{e1} 
V=-\epsilon\{P_2(\mathbf{m}_i \cdot \mathbf{m}_j)+\lambda [2(P_2 (\mathbf{e}_i \cdot \mathbf{e}_j)+
P_2 (\mathbf{e}_{\perp i} \cdot \mathbf{e}_{\perp j}))-P_2(\mathbf{m}_i \cdot \mathbf{m}_j)]\},
\end{equation}
where the strength parameter $\epsilon$ is a positive quantity setting energy and temperature scales (the dimensionless temperature used is defined as $T^*=k_BT/\epsilon$); 
$\lambda$ is a shape anisotropy of the biaxial molecular polarizability, usually termed as the biaxiality parameter. As stated in Ref. \cite{son} for the interaction potential Eq. (\ref{e1}), depending upon different mechanisms of molecular alignment, there can be two ranges of the biaxiality parameter:   $0\le \lambda \le \frac{1}{3}$ and $\frac{1}{3}<\lambda \le 1$. We investigated the case for which $0\le \lambda \le \frac{1}{3}$. 

\indent The pair potential can also be written in terms of symmetry-adapted ($D_{2h}$) Wigner functions $R_{mn}^L$, 
\begin{equation}\label{e2} 
V=-\epsilon\{R_{00}^2(\Omega_{ij})+6\lambda R_{22}^2(\Omega_{ij})\},
\end{equation}
Here $\Omega_{ij}=\{\phi_{ij},\theta_{ij},\psi_{ij}\}$ denotes the triplet 
of Euler angles defining the relative orientation of $i^{th}$ 
and $j^{th}$ molecules. To define the Euler angles, we have followed the convention used by Rose \cite{rose}. 
Thus, the Wigner functions $R_{mn}^L$ symmetrized for the $D_{2h}$ group of the $N_B$ phase are
 
\begin{equation}\label{e2}
R_{00}^2=\frac{3}{2}\cos^2\theta-\frac{1}{2},
\end{equation}  
\begin{equation}\label{e3}
R_{02}^2=\frac{\sqrt{6}}{4}\sin^2\theta\cos2\psi,
\end{equation}  
\begin{equation}\label{e4}
R_{20}^2=\frac{\sqrt{6}}{4}\sin^2\theta\cos2\phi,
\end{equation}  
\begin{equation}\label{e5}
R_{22}^2=\frac{1}{4}(1+\cos^2\theta)\cos2\phi\cos2\psi-\frac{1}{2}\cos\theta\sin2\phi\sin2\psi.
\end{equation}  

The above simple model reproduces both 
the uniaxial and the biaxial orientational orders \cite{son}. 
Besides the prediction of a direct $I$-$N_B$ transition for a range of values of the molecular biaxiality 
parameter $\lambda$, the MF theory predicts the existence of a tricritical point on the $N_U$ - $N_B$ transition line in the resulting phase diagram. In the MF study \cite{son}, the $N_B$ - $N_U$ transition is of second-order for $0<\lambda \lesssim 0.20$  and of first-order for $0.20<\lambda \lesssim 0.22$. The MF analysis thus predicts a tricritical point in the transition from uniaxial to biaxial at $({T^*}_t,\lambda_t)\approx(1.273,0.20)$ and a triple point at 
$({T^*}_c,\lambda_c)\approx(1.322,0.22)$ in the $\lambda - T^*$ plane. 

\indent In our simulations we consider a range of values of the biaxiality parameter starting from a lower value of $\lambda=0.14$ upto a maximum value of $\lambda=0.33$. This range of $\lambda$ values essentially contains all the important features of the phase diagram to be studied. To explore the phase diagram we carried out a series of MC simulations using the conventional Metropolis algorithm on a periodically repeated simple cubic lattice, consisting of   
$N=40^3$ particles. Simulations were run in 
cascade in order of increasing dimensionless temperature $T^*$. Equilibrium runs were between $30~0000$ and $500~000$ cycles and were followed by a production run of $500~000$-$800~000$ cycles (longer runs were used close to the transitions). 
An orientational move was attempted following 
the Barker-Watts method \cite{bar}.     
For a given  value of $\lambda$ the simulation at
the lowest temperature studied was started from the 
perfectly ordered state. 
The simulations at the other temperatures for the same $\lambda$ are run in 
cascade starting from an equilibrium configuration at a nearby lower 
temperature.

\indent To analyze the orientational order we computed the second rank order
parameters $\langle R_{mn}^2\rangle$ following the procedure prescribed
by Vieillard-Baron \cite{vie}. 
According to this study, a $\bf{Q}$ 
tensor is defined for the molecular axes associated with a 
reference molecule. For an arbitrary unit vector {$\bf{w}$}, 
the elements $\bf{Q}$ are defined as 
$Q_{\alpha\beta}(\bf{w}) = \langle(3 w_\alpha w_\beta-\delta_{\alpha\beta})/2\rangle$,
where the average is taken over the configurations and the subscripts $\alpha$ 
and $\beta$ label Cartesian components of $\bf{w}$ relative to an 
arbitrary laboratory frame. Diagonalizing this matrix one obtains nine 
eigenvalues and nine eigenvectors. These are then recombined to give 
the four order parameters 
$\langle R_{00}^2\rangle$, $\langle R_{02}^2\rangle$, 
$\langle R_{20}^2\rangle$ and $\langle R_{22}^2\rangle$  with respect to 
the directors. 
Out of these four second rank order parameters 
the uniaxial order parameter $\langle R_{00}^2\rangle$ and the biaxial order parameter $\langle R_{22}^2\rangle$ are involved in our study. The uniaxial order parameter $\langle R_{00}^2\rangle$ measures the alignment of longest molecular symmetry axis ($\mathbf{m}$) with respect to the laboratory $Z$ axis, conventionally chosen parallel to the primary director ($\mathbf{n}$). On the other hand $\langle R_{22}^2\rangle$ measures the alignment of two short transverse axes ($\mathbf{e}$, $\mathbf{e}_\perp$) along the laboratory $X$ and $Y$ axes and this order parameter identifies the phase biaxiality formed by biaxial molecules \cite{camp}. 

\indent The other physical quantities of interest in the present study are the response functions, the heat capcity $C_V$ and the ordering susceptibility $\chi_{R_{mn}^2}$, calculated respectively from the fluctuations in the energy and in the order parameter:
\begin{equation}\label{e12}
C_V = \frac{\langle {E}^2 \rangle-{\langle E \rangle}^2}{L^3{T}^2}; 
\end{equation}
\begin{equation}\label{e13}
\chi_{R_{mn}^2} = \frac{L^3(\langle {R_{mn}^2}^2 \rangle-{\langle R_{mn}^2 \rangle}^2)}{T}.
\end{equation}
where $E$ is the scaled total energy ($E=\mathcal{E}/\epsilon$) of the system. 

\section{RESULTS} 
\begin{figure}[tbh]
\begin{center}
\resizebox{120mm}{!}{\rotatebox{270}{\includegraphics[scale=1.0]{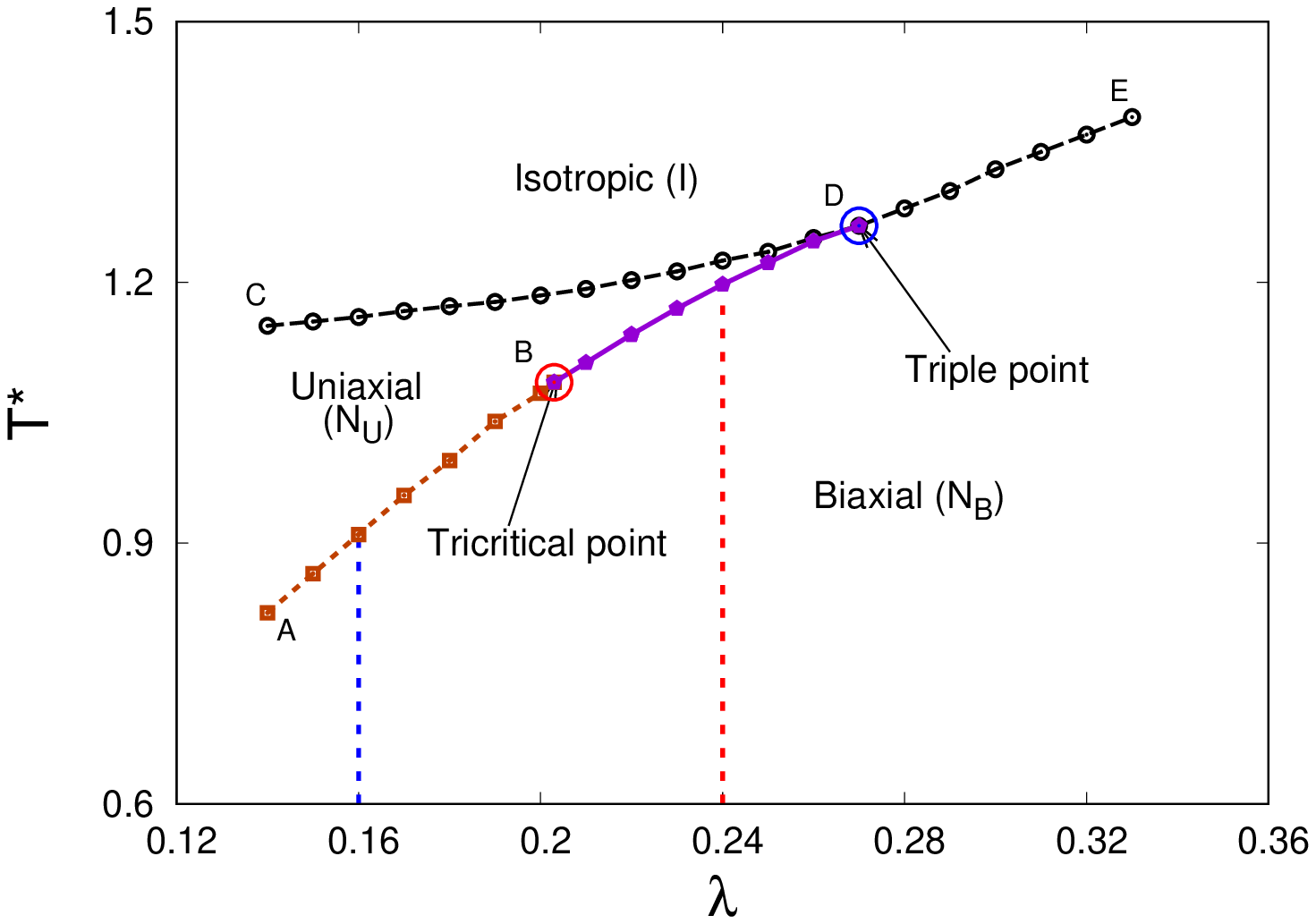}}}
\end{center}
\caption{The phase diagram with a tricritical point ($B$)  and a triple point ($D$) in the $\lambda - T^*$ plane. MC results are shown as points and the lines joining them are used as a guide to the eye. The broken line represents the second-order phase transitions while the solid lines represent the first-order transitions. The vertical dotted lines mark the points on the $N_U$-$N_B$ coexistence line at which extensive simulations are performed to locate the tricritical point.}
\label{f1}
\end{figure}

\indent We first present the MC phase diagram in Fig.\ref{f1} showing the variation of the transition temperatures ($T^*$) with the biaxiality parameter ($\lambda$) for the biaxial model computed for $0< \lambda <\frac{1}{3}$ and for a cubic lattice of size $L=40$. 
The phases were identified by investigating the orientational order parameters $\langle R_{00}^2\rangle$ and $\langle R_{22}^2\rangle$. The transition temperatures (Table~\ref{tab:t1}) were evaluated from the peaks in the temperature variation of the respective order parameter susceptibilities.
In the phase diagram (Fig.\ref{f1}), the broken line $AB$ represents the second order phase transitions while the solid line $BD$ represents the first order transitions between the uniaxial and biaxial phases. The point $B$ where these two lines meet is the tricritical point ($\lambda_t$, $T^*_t$). The  solid line $CD$ represents the first order transitions between isotropic and uniaxial phases. The triple point  where the three phases coexist in the phase diagram is denoted by $D$ with coordinates ($\lambda_c$, $T^*_c$). 
Beyond $\lambda > \lambda_c$ the direct $I$ to $N_B$ transitions take place across the coexistence line $DE$ having a first-order character in our case.  

\begin{table}[h]
\caption{\label{tab:t1} MC estimates for the transition temperatures, obtained from $\chi_{R^2_{mn}}~{\it vs}~T^*$ plots for different values of the biaxiality parameter $\lambda$, and based on the largest simulated lattice size $L=40$. The error in each temperature is within $\pm 0.0025$.}
\begin{center}
\begin{tabular}{ c c c c}
\hline
\hline 
$\lambda$ &  $T^*_{IU}$ & $T^*_{UB}$ & $T^*_{IB}$\\
\hline
 0.14 &1.1500 &0.8200 &  \\
 0.15 &1.1550 &0.8650 &  \\
 0.16 &1.1600 &0.9125 &  \\
0.17 &1.1670 &0.9550 &  \\
 0.18 &1.1725 &0.9950 &  \\
0.19 &1.1775 &1.0400 &  \\
0.20 &1.1850 &1.0725 &  \\
0.21 &1.1925 &1.1075 & \\
0.22 &1.2025 &1.1400 &  \\
0.23 &1.2125 & 1.1700&  \\
0.24 &1.2250 &1.1975 &  \\
0.25 &1.2350 &1.2225 &  \\
0.26 &1.2475 &1.2440 &  \\
0.27 &1.2635 &1.2635 &  \\
0.28 & & &1.2850  \\
0.29 & & &1.3050   \\
0.30 & & &1.3300   \\
0.31 & & & 1.3500\\
0.32 & & &1.3700  \\
0.33 & & & 1.3900 \\
\hline
\hline 
\end{tabular}
\end{center}
\end{table}

\begin{figure}[tbh]
\begin{center}
\resizebox{120mm}{!}{\rotatebox{270}{\includegraphics[scale=1.0]{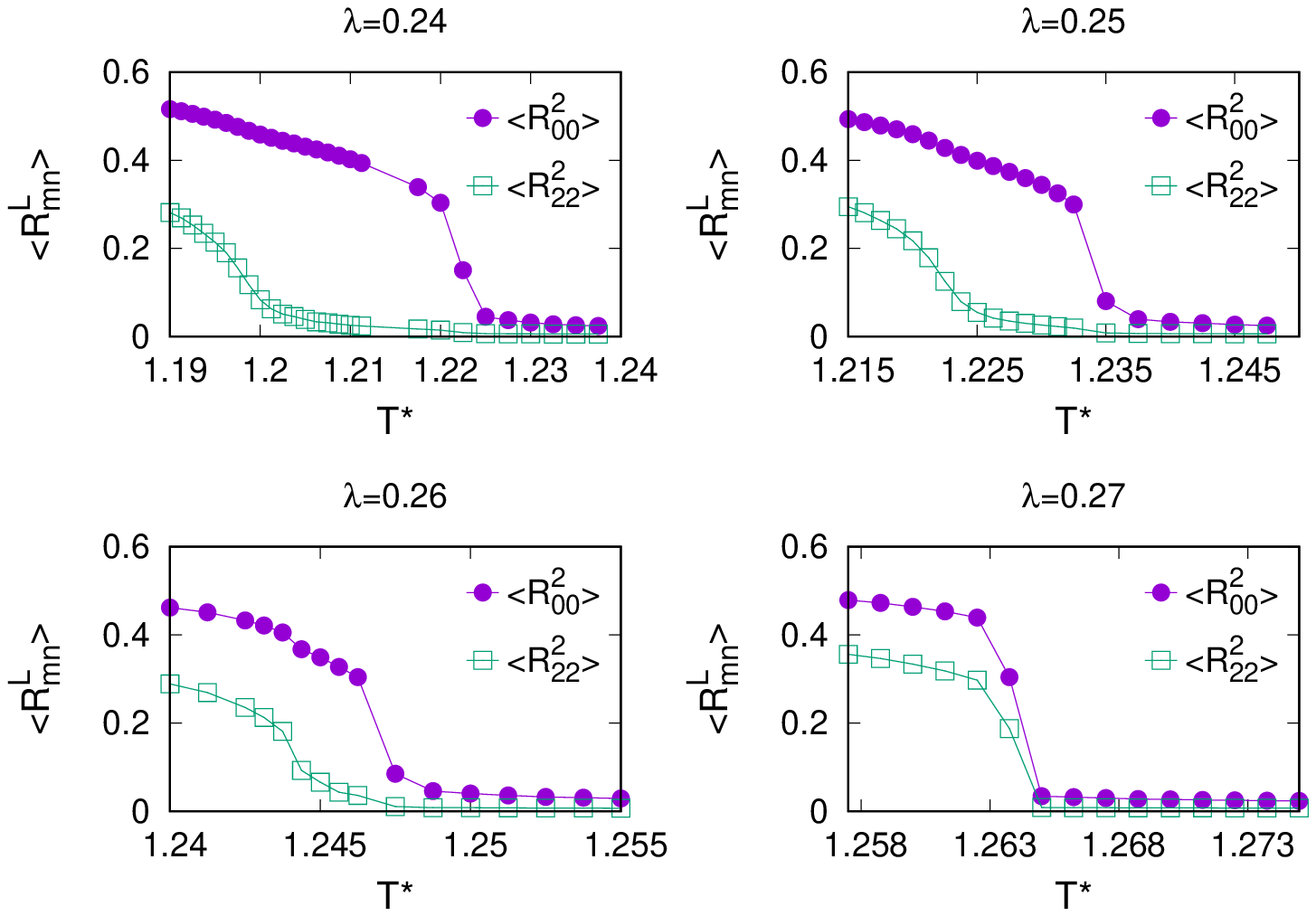}}}
\end{center}
\caption{The second-rank orientational order parameters ($\langle R_{mn}^L\rangle$) versus dimensionless temperature ($T^*$) for the molecular biaxiality parameter $\lambda = 0.24, 0.25, 0.26, 0.27$ obtained from MC simulations for the lattice size $L=40$. The associated statistical errors are within point size.}
\label{f2}
\end{figure}

\indent To locate the triple point on the phase diagram first we investigated the temperature variation of the orientational order parameters. In Fig. \ref{f2}, MC results for $\langle R_{00}^2\rangle$ and $\langle R_{22}^2\rangle$ are shown for $\lambda = 0.24, 0.25, 0.26, 0.27$. Both the order parameters decrease monotonically with reduced temperature. Starting with the lower values of $\lambda$ we observe that the uniaxial order parameter $\langle R_{00}^2\rangle$ appears to vanish at a higher temperature indicating the usual first-order $N_U-I$ transition, while the biaxial order parameter $\langle R_{22}^2\rangle$ vanishes at a lower temperature indicating the $N_B-N_U$ transition. For $\lambda=0.26$ both the transitions $N_U-I$ and $N_B-N_U$ get shifted towards higher temperature and simultaneously become closer. They transform into a single direct $N_B-I$ transition for $\lambda=0.27$ at a higher temperature of  $T^*_{BI}=1.2650$. This observation gives a clear indication of the occurrence of the triple point in the phase diagram. 

\begin{figure}[tbh]
\begin{center}
\resizebox{120mm}{!}{\rotatebox{270}{\includegraphics[scale=0.8]{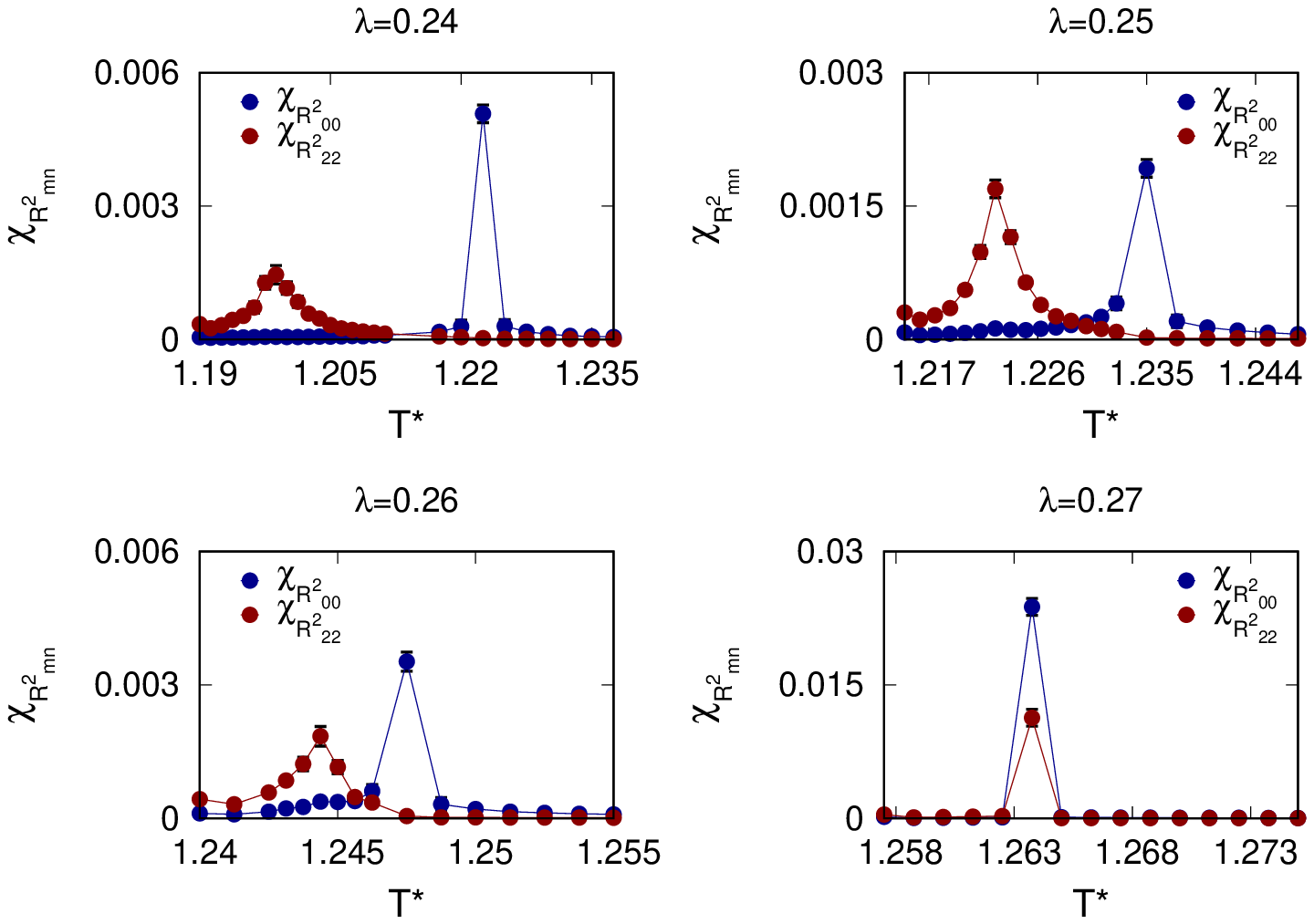}}}
\end{center}
\caption{The susceptibilities ($\chi_{R^2_{mn}}$) for the uniaxial and biaxial order parameters versus dimensionless temperature ($T^*$) for $\lambda = 0.24, 0.25, 0.26, 0.27$ obtained from MC simulations for the lattice size $L=40$. The associated statistical errors are shown by error bars.}
\label{f3}
\end{figure}
\indent The corresponding susceptibilities $\chi_{R^2_{00}}$ and $\chi_{R^2_{22}}$ are shown in Fig.~\ref{f3}. For $\lambda = 0.24$, the peak of $\chi_{R^2_{22}}$, that corresponds to the $N_B - N_U$ transition, occurs at a low temperature ($T^*_{BU}=1.1975$), while the peak of $\chi_{R^2_{00}}$ associated to the $N_U - I$ transition occurs at a higher temperature ($T^*_{UI}=1.2250$). As the biaxiality increases the two peaks approach one another and for $\lambda = 0.27$ they coalesce into a single one at a dimensionless temperature of $1.2635$. Thus, the MC estimate for the triple point is $(T^*_c,\lambda_c) \approx (1.2635, 0.27)$. This differs significantly from the mean field prediction reported in Ref. \cite{son}. The shifting of triple point towards higher $\lambda$ was previously reported by Romano in Ref.\cite{rom1}. However, the position of the triple point was not precisely determined in Ref.\cite{rom1}.   

\begin{figure}[tbh]
\begin{center}
\resizebox{120mm}{!}{\rotatebox{270}{\includegraphics[scale=1.0]{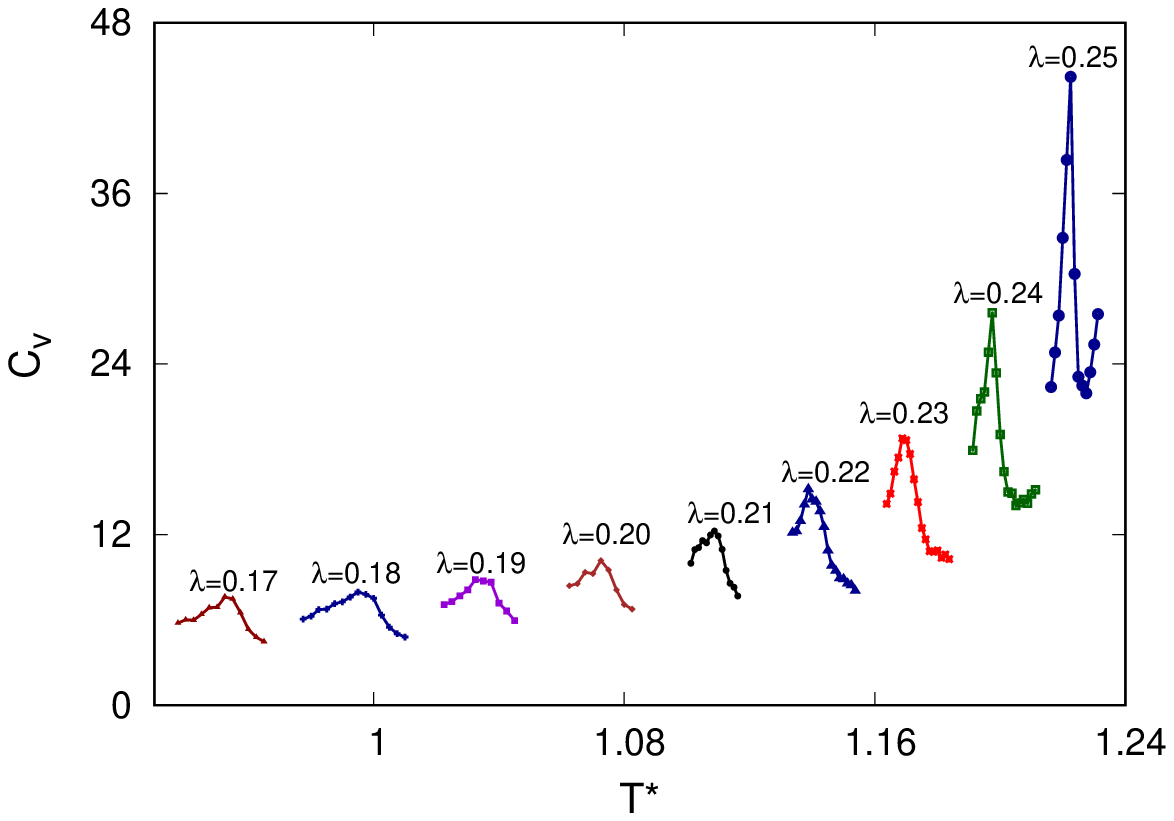}}}
\end{center}
\caption{ Variation of configurational heat capacity with reduced temperature for nine different degrees of molecular biaxiality $\lambda=0.17, 0.18, 0.19, 0.20, 0.21, 0.22, 0.23, 0.24$, and $0.25$. MC results, obtained for the lattice size $L=40$, are shown as different symbols for different biaxialities.}
\label{f4}
\end{figure}

\indent Next we investigated the tricritical point by calculating the internal energy, the uniaxial and biaxial order parameters and their corresponding response functions. The variation of configurational heat capacity $C_V$ with dimensionless temperature $T^*$, in the vicinity of the uniaxial-biaxial transition for different biaxialities, is shown in Fig.~\ref{f4}. With increasing biaxiality the peak height of the heat capacity curve increases. Distinct peak in $C_V$ appears for $\lambda\gtrsim0.22$ indicating approximately the location of crossover from second to first order transitions.

\begin{figure}[tbh]
\begin{center}
	\resizebox{120mm}{!}{\rotatebox{270}{\includegraphics[scale=1.0]{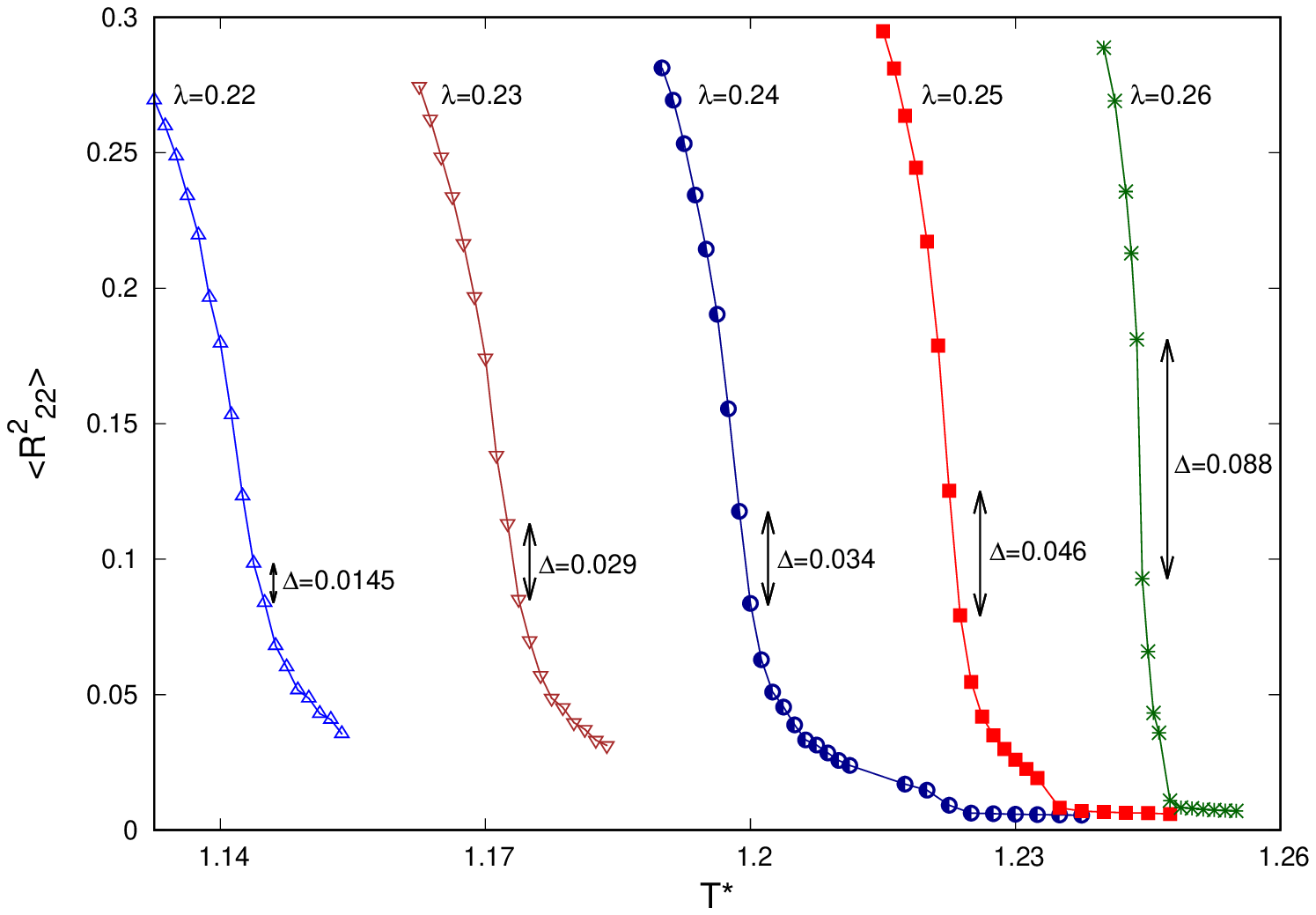}}}
\end{center}
\caption{Variation of the biaxial order parameter with reduced temperature for five different values of the biaxiality parameter $\lambda = 0.22, 0.23, 0.24, 0.25, 0.26$. As the biaxiality parameter increases the jump ($\Delta$) in $\langle R^2_{22}\rangle$ at the first-order uniaxial-biaxial transition become more pronounced.}
\label{f5}
\end{figure}

\indent Simulation results for the biaxial order parameter ($\langle R_{22}^2\rangle$) are shown in Fig.~\ref{f5} for five different values of the biaxiality parameter $\lambda = 0.22, 0.23, 0.24, 0.25, 0.26$. The order parameter $\langle R^2_{22}\rangle$ shows a discontinuous jump at the uniaxial-biaxial transition and for the higher values of $\lambda$ the jump becomes more pronounced. This behavior indicates that a change in character of the uniaxial-biaxial transition occurs in the vicinity of $\lambda \approx 0.22$, yielding there a tricritical point on the $N_U$ - $N_B$ transition line. 

\begin{figure}[tbh]
	\begin{center}
		\resizebox{120mm}{!}{\rotatebox{270}{\includegraphics[scale=1.0]{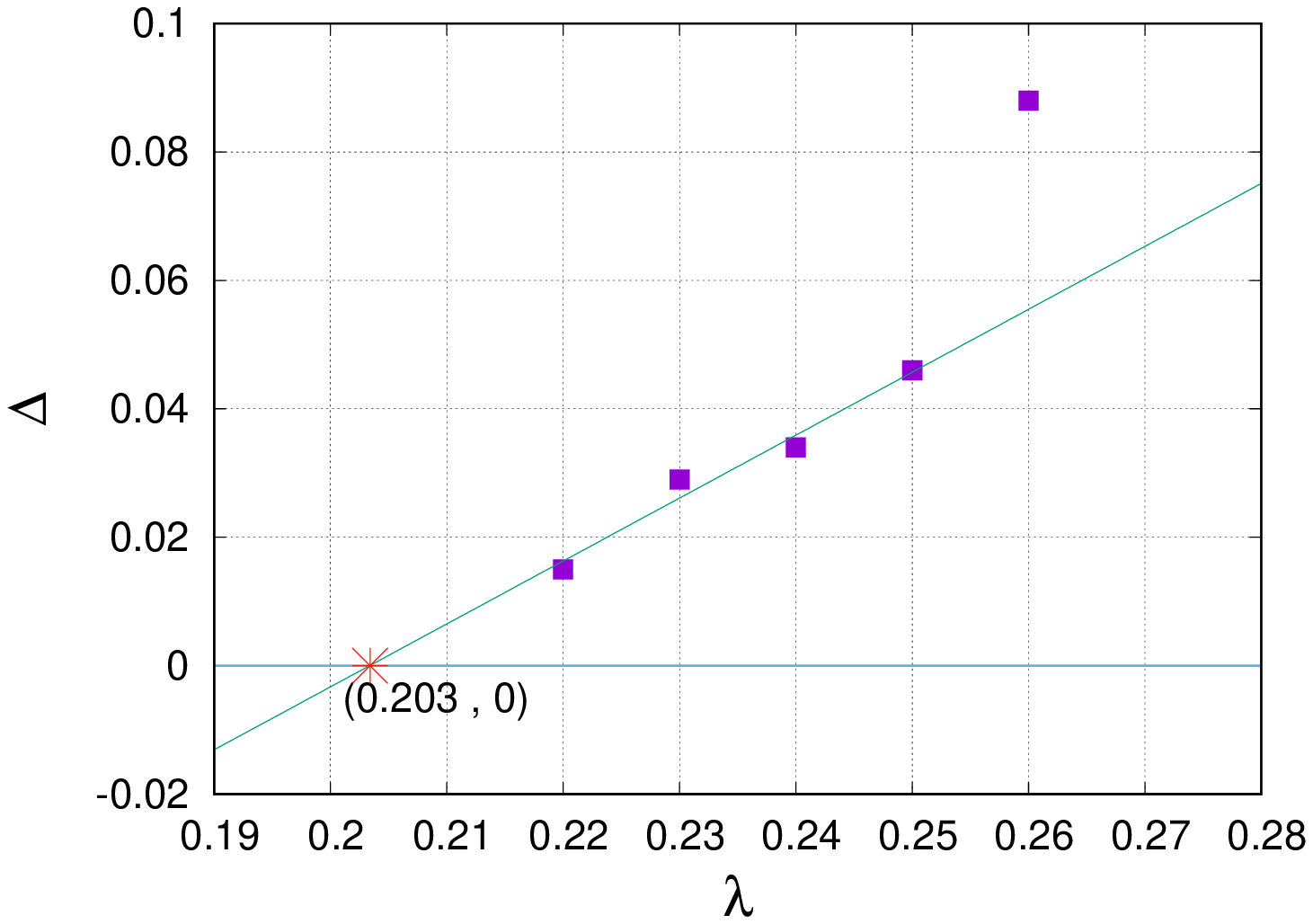}}}
	\end{center}
	\caption{Linear fit of the jump ($\Delta$) in $\langle R^2_{22}\rangle$ using four values of the biaxiality parameter $\lambda = 0.22, 0.23, 0.24, 0.25$. The extrapolation provides an estimate of $\lambda_t$ by extending the linear fit up to $\Delta=0$ and yields $\lambda_t \approx 0.203$. The point corresponding to $\lambda=0.26$ is excluded from the linear fit due to large non-linear deviation.}
	\label{f6}
\end{figure}

\indent To estimate the value of $\lambda$ that corresponds to the tricritical point, we used an extrapolation to the decreasing values of $\Delta$ with lowering $\lambda$. 
As shown in Fig.~\ref{f6}, for lower $\lambda$ values, the jump in $\langle R^2_{22}\rangle$, i.e., $\Delta$, varies linearly with $\lambda$. The extrapolated value of $\lambda$, obtained from the linear fit, is $\lambda_t \approx 0.203$ and this is very close to the theoretical value reported in Ref. \cite{son}.   

\begin{figure}[tbh]
	\begin{center}
		\resizebox{120mm}{!}{\rotatebox{270}{\includegraphics[scale=1.0]{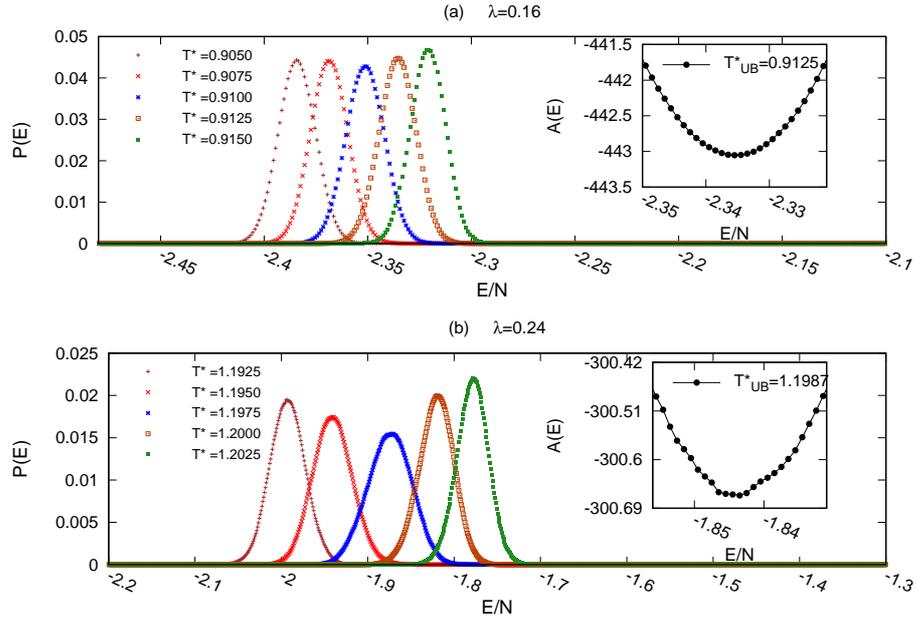}}}
	\end{center}
	\caption{Probability distributions obtained from MC simulations for different temperatures in the vicinity of the transition with (inset) the associated free energy $A(E)$ as a function of energy per particle for the lattice size $L=40$: (a) $\lambda = 0.16$; (b) $\lambda = 0.24$.}
	\label{f7}
\end{figure}

\indent Additionally, to confirm the existence of the tricritical point on the $N_U$ - $N_B$ transition line, we investigated the behavior of a free-energy-like function at the transition for two $\lambda$-values, one above and the other below the mean field value at the same distance of it, namely, $\lambda = 0.2\pm0.04$. We employed the multiple histogram reweighting technique \cite{new} to calculate the relevant part of the free-energy-like functions $A(E)$ using the relation $A(E)=-\ln P(E)$. Here,  $P(E)=h(E)/\sum_E h(E)$ represents the normalized histogram. In a first-order transition, the free energy displays a double-well structure near the transition; in contrast, the free energy has a single minimum for a second-order transition \cite{lee1,lee2}.   
For $\lambda=0.16$ there is only one minimum in $A(E)$ (Fig.~\ref{f7}(a)) demonstrating a second-order transition. By analyzing the Probability distributions $P(E)$ and the derived free-energy-like function $A(E)$ for $\lambda=0.24$, we observe that there is very little evidence of a barrier forming for the lattice size $L=40$. Expecting that a double-well structure in the free energy might be produced for higher lattice sizes at this biaxiality, we have generated histograms near the uniaxial-biaxial transition with a higher lattice size of $L=60$. As expected, for this higher system size, a sign of a barrier in $A(E)$ is obtained indicating the first-order nature of the $N_U$ - $N_B$ transition for the biaxiality $\lambda=0.24$.  
  
\begin{figure}[tbh]
	\begin{center}
		\resizebox{120mm}{!}{\rotatebox{270}{\includegraphics[scale=1.0]{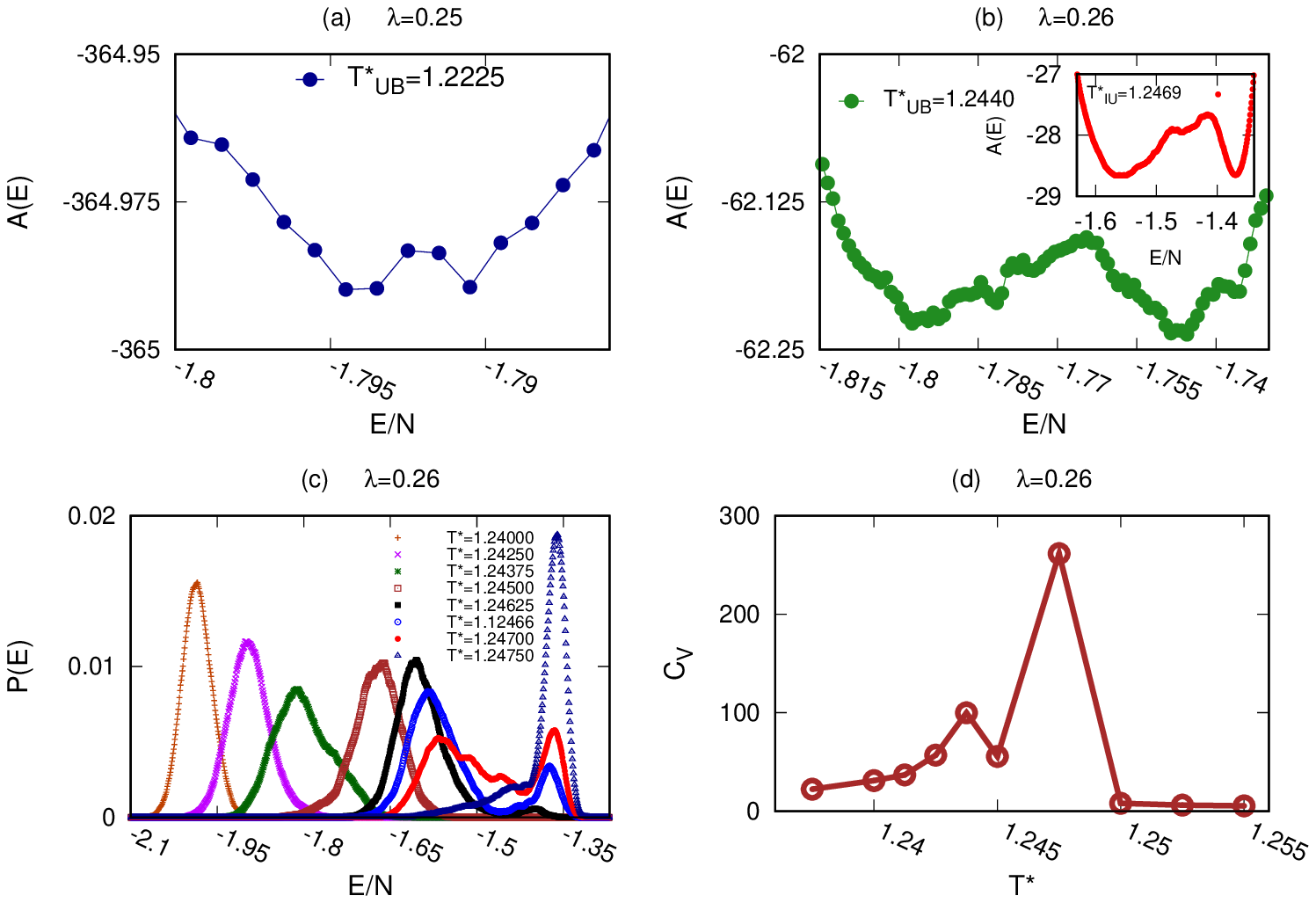}}}
	\end{center}
	\caption{(a) Free energy $A(E)$ as a function of energy per particle at $T=T^*_{UB}$ for $\lambda = 0.25$; (b) free energy $A(E)$ as a function of energy per particle at $T=T^*_{UB}$ with (inset) the same at $T=T^*_{IU}$ for $\lambda = 0.26$; (c) probability distributions obtained from MC simulations for different temperatures in the vicinity of both the transitions $I-N_U$ and $N_U-N_B$ for $\lambda = 0.26$; (d) heat capacity $C_V$ as a function of reduced temperature $T^*$ for $\lambda = 0.26$ and $L=40$.}
	\label{f8}
\end{figure}

\indent In order to provide conclusive evidence of the $BD$-branch's first-order character in the phase diagram (Fig.~\ref{f1}), two further simulations were conducted for two higher $\lambda$ values, $0.25$ and $0.26$. As shown in Fig.~\ref{f8}(a),(b), a distinct double-well structure forms in $A(E)$ for both of these higher $\lambda$ values.      
The simulation results for the free-energy-like function $A(E)$ thus provide a conclusive numerical evidence of the existence of a tricritical point on the $N_U$ - $N_B$ transition line within the interval $0.16<\lambda<0.24$. 

\indent The inset of Fig.~\ref{f8}(b), showing the variation of $A(E)$ versus $E$ for the $I$-$N_U$ transition generated at the reduced temperature $T^*_{IU}=1.2469$, is reported to compare the degrees of first-orderedness with the $N_U$-$N_B$ transition for the same value of the biaxiality $\lambda=0.26$. For the comparison we define a quantity $\Delta A = A(E_m)-A(E_1)$, where $E_1$ is the energy at which the two minima of $A$ of equal depth appear and $E_m$ denotes the energy of the maximum of $A$ between the two minima. $\Delta A$ is a measure of the bulk free energy barrier of the system \cite{lee1,lee2}. In the inset of Fig.~\ref{f8}(b) it is seen that the free-energy barrier $\Delta A$ for the $I$-$N_U$ transition is almost ten times greater than that for the $N_U$-$N_B$ transition at $\lambda=0.26$ and for the system size $L=40$. This observation establishes the fact that the first-order $N_U$-$N_B$ transitions on the coexistence line $BD$ in phase diagram (Fig.~\ref{f1}) are much weaker than the usual weak first-order isotropic-uniaxial transition \cite{zha}. 

\indent For $\lambda = 0.26$, the energy histograms for different temperatures in the vicinity of both the transitions $I-N_U$ and $N_U-N_B$ are presented in Fig.~\ref{f8}(c). It is evident from the histogram plots that the $N_U-N_B$ transition closely follows the $I-N_U$ transition, as is stated earlier, indicating the shift of triple point towards higher $\lambda$. The free energies in Fig.~\ref{f8}(b) were derived using these histograms. The dual peaks in the heat capacity ($C_V$) (Fig.~\ref{f8}(d)) correspond the two successive transitions at $T^*_{IU}=1.2475$ and $T^*_{UB}=1.2440$ respectively for $\lambda = 0.26$.   


\section{CONCLUSION}
In conclusion, we present the Monte Carlo phase diagram of a lattice model of biaxial nematogenic molecules interacting with the Straley's quadrupolar pair potential in Sonnet-Virga-Durand parameterization that gives both the uniaxial and the biaxial nematic phases along with a tricritical point in the transition from uniaxial to biaxial nematics. Multiple histogram reweighting technique is employed to derive a free-energy-like function and it is used for analyzing the order of phase transition. Our study thus, provides a conclusive numerical evidence in support of the existence of a tricritical point on the uniaxial-biaxial transition line in the phase diagram. Another important finding of our investigation is that the first-order uniaxial-biaxial transition is much weaker than the usual weak first-order isotropic-uniaxial transition.

\indent We conclude that although the qualitative nature of the MC phase diagram is identical as obtained in theoretical study however the location of the triple point differs significantly from theoretical prediction.  
\section{ACKNOWLEDGMENT}
\indent We are grateful to Soumen Kumar Roy (Jadavpur University, Kolkata, India)  for useful discussions. S.DG. acknowledges support through a research
grant obtained from Council of Scientific and Industrial
Research (03/1235/12/EMR-II).\\

\end{document}